\documentclass[RNAAS]{aastex62}

\usepackage{graphicx}
\usepackage{natbib}
\usepackage{latexsym}
\usepackage{amssymb}
\usepackage{longtable}
\usepackage{amsmath}
\usepackage{url}
\begin{document}

\title{No consistent atmospheric absorption detected for the ultra-hot Jupiter WASP-189 b}

\correspondingauthor{P. Wilson Cauley}
\email{pwcauley@gmail.com}

\author{P. Wilson Cauley}
\affiliation{Laboratory for Atmospheric and Space Physics, University of Colorado Boulder, Boulder, CO 80303}

\author{Evgenya L. Shkolnik}
\affiliation{Arizona State University, School of Earth and Space Exploration, Tempe, AZ 85287}

\author{Ilya Ilyin}
\affiliation{Leibniz-Institute for Astrophysics Potsdam (AIP), An der Sternwarte 16, 14482, Potsdam, Germany}

\author{Klaus G. Strassmeier}
\affiliation{Leibniz-Institute for Astrophysics Potsdam (AIP), An der Sternwarte 16, 14482, Potsdam, Germany}

\author{Seth Redfield}
\affiliation{Wesleyan University, Astronomy Department, Van Vleck Observatory, Middletown, CT}

\author{Adam Jensen}
\affiliation{Department of Physics and Astronomy, University of Nebraska at Kearney, Kearney, NE 68849}


\section{} 

Transiting ultra-hot Jupiters (UHJs), which have day-side temperatures of
$\gtrsim 2200$ K \citep{parmentier18}, offer the opportunity to explore an
extreme regime of atmospheric physics. Peering through hot planet atmospheres
in transmission at high spectral resolution can provide details on the
composition \citep{hoeijmakers18,casasayas19,keles19}, temperature structure
\citep{wyttenbach15,wyttenbach17}, geometry \citep{yan18}, and dynamics
\citep{louden15,allart18,cauley19} of material in the extended
gravitationally-bound thermosphere. In this research note we report on the null
detection of an atmospheric signature around the ultra-hot Jupiter WASP-189 b
\citep{anderson18}.  WASP-189 b orbits a bright ($V=6.64$) A6IV-V star with an
orbital period of $\approx 2.7$ days and is inflated with a radius of
$R_\text{p} \approx 1.4 R_J$. Its day-side equilibrium temperature of $\approx
2640$ K makes it one of the hottest gas giants discovered to-date.

We observed the transit of WASP-189 b between 05:15--10:50\,UT on May 15, 2019
with the Large Binocular Telescope in Arizona and its high-resolution \'echelle
spectrograph PEPSI \citep{strassmeier15}. Due to poor weather conditions, the
first half of the transit was not observed and no pre-transit observations were
obtained.  Clouds also prevented continuous post-transit exposures, resulting
in a gap between 07:15 UT and 09:05 UT. PEPSI was used in its $R \approx
50,000$ mode and with cross dispersers (CD) III (blue arm) and V (red arm)
simultaneously.  The wavelength coverage was 4750--5430\,\AA\ in the blue arm
and 6230--7430\,\AA\ in the red arm. The spectra were collected with a constant
signal-to-noise of 300 in the continuum controlled by a photon counter. All
data were reduced with the Spectroscopic Data System for PEPSI (SDS4PEPSI). The
observational setup and reduction procedures are identical to those outlined in
\citet{cauley19}. 

In addition to applying standard data reduction steps, we fit and removed
telluric absorption using \texttt{MOLECFIT} \citep{kausch15,smette15}. The
profile of the stellar surface occulted by the planet during transit was
removed using synthetic occulted surface profiles generated by a stellar
photosphere model and assuming the system parameters from \citet{anderson18}.
The model photosphere was created with \texttt{Spectroscopy Made Easy}
\citep{valenti96,piskunov17}. 

The transmission spectrum for each exposure is created by dividing each
spectrum by the signal-to-noise weighted mean out-of-transit spectra. The
individual transmission spectra are divided by the synthetic occulted surface
profiles to remove the effect of the occulted stellar surface. Each spectrum is
then shifted into the rest frame of the planet. The equivalent width of any
line of interest is then integrated across $\pm 50$ km s$^{-1}$ of the line's
rest wavelength.

\begin{figure}[b!]
\begin{center}

\includegraphics[scale=0.6,angle=0,clip,trim=0mm 10mm 0mm 20mm]{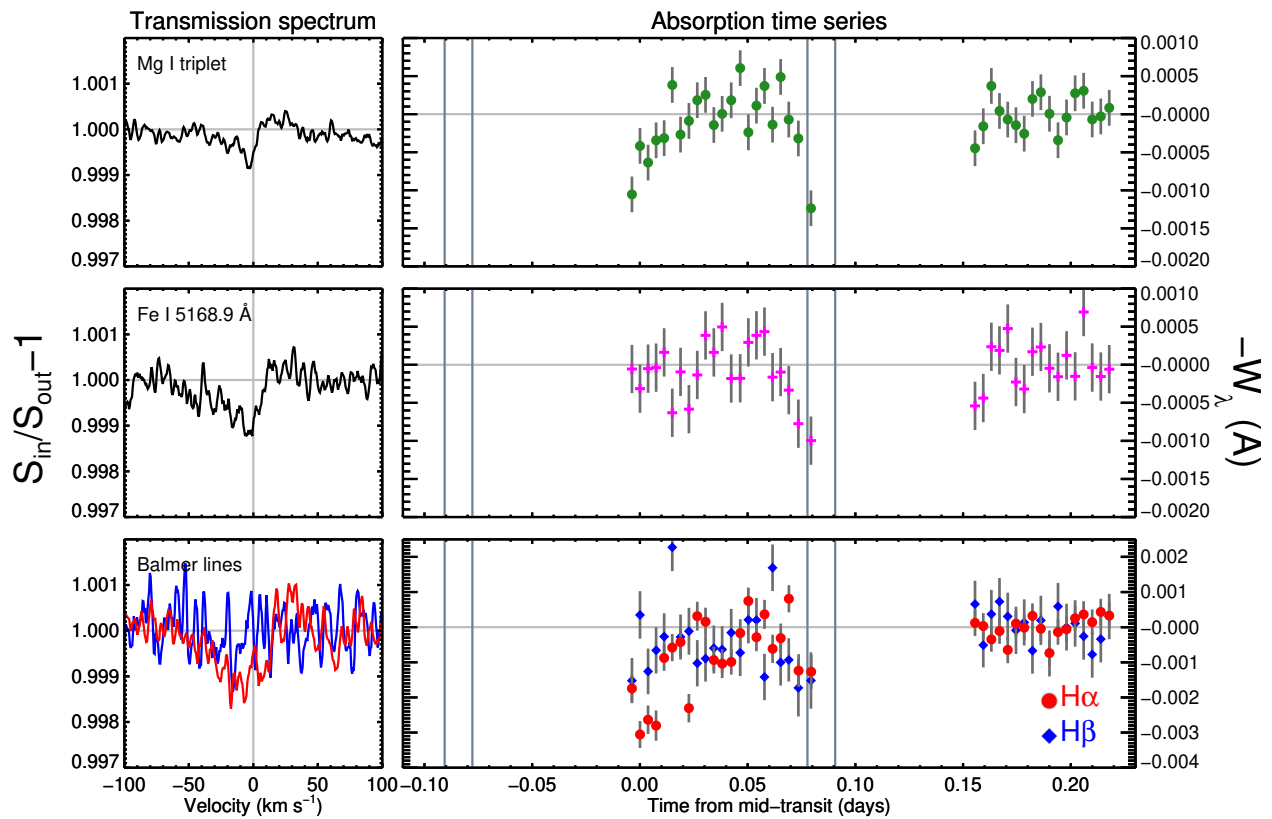}

\caption{Average transmission spectra (left panels) and time series absorption
(right panels) for the lines with detected absorption signatures. Note that the
\ion{Mg}{1} transmission spectrum is the average of all three triplet members.
The vertical gray lines in the time series panels mark the transit contact
points. Although there is some significant in-transit variability, no signal
consistent with an extended atmosphere around the planet is detect.
\label{fig:1}}

\end{center}
\end{figure}

We searched for absorption in a number of metal ions, including \ion{Fe}{1},
\ion{Fe}{2}, \ion{Mg}{1}, and \ion{Ti}{1} and in the Balmer lines H$\alpha$ and
H$\beta$. We found marginal ($\approx 1-2\sigma$) hints of absorption in the
mean transmission spectra of \ion{Mg}{1} triplet at 5167, 5173, and 5184 \AA\
and also in the \ion{Fe}{1} 5168.9 \AA\ line (see \autoref{fig:1}). However,
the absorption time series for these lines show no consistent in-transit
signal, suggesting that the mean transmission signals do not arise in the
planetary atmosphere. Similar marginal signals are seen in the Balmer lines
H$\alpha$ and H$\beta$. The time series absorption (right panels of
\autoref{fig:1}) is highly variable and the individual spectra with the
strongest absorption are likely driving the signal in the mean transmission
spectrum. 

Although we only observed a partial transit and lost segments of time to clouds,
the spectra we were able to collect are high signal-to-noise and thus we can
use the in-transit measurements to place constraints on the extent of the
planet's thermosphere. The depth of the H$\alpha$ transmission spectrum
places an upper limit on the radius of the excited hydrogen thermosphere of  
$\approx 0.2 R_\text{p}$ above the optical radius. The depth of the metal
line transmission spectra are even smaller, constraining their radial extent
to $< 0.2 R_\text{p}$. 

The lack of strong atomic transmission signatures in the atmosphere of WASP-189
b is surprising given its high temperature. Such absorption lines are easily
detectable in a number of other UHJs
\citep[e.g.,][]{jensen18,casasayas19,yan19,cauley19}. However, WASP-189 b has
the smallest transit depth of the known UHJs ($\approx 0.4\%$) decreasing the
depth of any transmission signatures for a given atmospheric radius compared to
the other UHJs. The highly variable nature of the in-transit signals suggests
that the transmission lines profiles are not the result of absorption in the
planet's atmosphere. We suggest that the in-transit features are instead due to
an inhomogeneous stellar disk and/or stellar variability caused by magnetic
surface features. Such features are now known to exist on A-type stars
\citep[e.g.,][]{petit17} and could produce the small variations that we observe
via the contrast effect \citep{cauley18}.

Our observations demonstrate the lack of a highly extended atmospheric around
WASP-189 b. However, more transit observations in better conditions and
covering the entire transit are needed to entirely rule out the presence of a
thermosphere above $\approx 1.2 R_\text{p}$. The depth of the features detailed
here should be taken into account for future transit observations in order to
place stronger constraints on the atmosphere than those we have presented.

\bibliography{references}{}
\bibliographystyle{aasjournal}

\end{document}